\documentclass[aps,prl,twocolumn,groupedaddress]{revtex4}
\usepackage{graphicx}
\usepackage{amsmath}

\begin{document}

\title{Error threshold estimates for surface code with loss of qubits}
\author{Masayuki Ohzeki}
\date{\today}

\address{Dipartimento di Fisica, Universit\`a di Roma `La Sapienza', P.le
Aldo Moro 2, 00185 Roma, Italy} \address{Department of Systems Science,
Graduate School of Informatics, Kyoto University, Yoshida-Honmachi,
Sakyo-ku, Kyoto, 606-8501, Japan}

\begin{abstract}
We estimate optimal thresholds for surface code in the presence of loss via an
analytical method developed in statistical physics. The optimal threshold for
the surface code is closely related to a special critical point in a
finite-dimensional spin glass, which is disordered magnetic material. We
compare our estimations to the heuristic numerical results reported in
earlier studies. Further application of our method to the depolarizing
channel, a natural generalization of the noise model, unveils its wider
robustness even with loss of qubits.
\end{abstract}

\maketitle

\textit{Introduction}--- Against corruption by environmental noise as well
as imperfection in implementation, the state of qubits describing quantum
information cannot be stable and must be recovered by elaborated procedures,
quantum error correction \cite{QEC1,QEC2}. Quantum error corrections usually
work on the computational error on qubits, which do not go out of the basis
for computations. Therefore errors come from losses of the physical
resource, qubits, can deprive of the performance of error correction.
However, if one can detect and identify locations of losses, a modified
scheme can recover the original information. Stace and Barret have suggested
an error-correcting code, a family of Kitaev's surface codes \cite%
{Dennis}, which is robust against both of the computational errors and
losses by a modified scheme according to the location of the lost qubits 
\cite{Loss1}.

In the present paper, we estimate precise values of the error thresholds for
the modified error correcting code against both of the computational errors
and loss of qubits by use of a systematic theory developed in statistical
physics. The key of our analysis is hidden in the disordered magnetic
system, spin glasses. Several spin glass models have a special symmetry with
exact solvable subspace known as Nishimori line \cite{HNbook,HN81}. The
critical point in this subspace, termed as the multicritical point,
corresponds to the optimal error threshold in the surface code \cite{Dennis}. A combination of the duality with the
real-space renormalization technique, which are often used to identify the
singular points in statistical mechanical models, can derive the precise
estimations for the optimal error thresholds \cite{NN,MNN} and
systematically approach the exact solutions \cite{ONB,Ohzeki}. By use of
this method, we fill the blank on the analytical study for the optimal error
thresholds on several surface codes with loss of
qubits. The results reported in this paper provide upper bounds against
error rates for any error-correcting schemes. They serve as important
benchmarks with which any constructive error correcting procedure as
recently proposed in Ref. \cite{HM} can be compared.

\textit{Surface code and spin glass}--- Let us consider qubits set on each
edge $(ij)$ of the square lattice embedded on a torus (genus $1$). We define
the star operator $X_s = \otimes_{(ij) \in s}X_{(ij)}$ for each site $s$,
and plaquette operator $Z_p = \otimes_{(ij) \in p}Z_{(ij)}$ for each
plaquette $p$ (site on the dual lattice), where $X$ and $Z$ are Pauli
matrices. The product consists of four edges adjacent to each site or
plaquette. The stabilizer group is given by the simultaneous eigenstates
with the positive eigenvalues for these operators $X_s$ and $Z_p$. Since the
star and plaquette operators consist of unit loops on the dual and original
square lattices, any contractible loop by $X_s$ and $Z_p$ products on each
lattice acts trivially on the codespace. On the other hand, any
non-contractible loops on the lattice can map the codespace to itself in a
nontrivial manner. If we set $L\times L$ lattice on a torus, we have $2L^2$
qubits and $2(L^2-1)$ stabilizers. The remaining degrees of freedom of $2$
implies existence of two non-contractible loops, winding around the hole of
the torus $L_v$ and winding around the body of the torus $L_t$, and ones $%
L_v^*$ and $L_t^*$ on the dual lattice. These loops can be written in terms
of the products of operators as $\bar{Z}_v = \prod_{(ij) \in L_v}Z_{(ij)}$, $%
\bar{X}_v = \prod_{(ij) \in L_v^*}X_{(ij)}$, $\bar{Z}_t$, and $\bar{X}_t$,
which are termed as logical operators. The logical operators can form Pauli
algebra of two effective qubits encoded in the topological degrees of
freedom on the torus as $[\bar{Z}_v,\bar{Z}_t]=[\bar{X}_v,\bar{X}_t]=0$, and 
$\bar{X}_t\bar{Z}_v = (-1)^{\delta_{tv}}\bar{Z}_v\bar{X}_t$. The
combinations of non-contractible loops yield $2^4 = 16$ different homology
classes for the original and dual square lattices on a single torus. We need
to distinguish them for protecting the information from corruption.

In order to evaluate the performance of the error-correcting code, let us
define a noise model where each qubit independently gets errors as 
\begin{equation}
\rho \to p_I\rho + \left(p_XX\rho X + p_YY\rho Y + p_ZZ\rho Z\right).
\end{equation}
Although, if we employ the following analytical method, we can estimate
precise values of the error thresholds for ``any" cases of $p_I,p_X,p_Y$,
and $p_Z$, we restrict ourselves to two cases: $p_X=p_Z=p$, $p_Y=p^2$ and $%
p_I=(1-p)^2$ (uncorrelated case), and $p_X=p_Y=p_Z=p/3$ and $p_I=1-p$
(depolarizing channel case) for simplicity, where $0 \le p \le 1$. The error 
$Y_{(ij)}$ can be regarded as a multiple error $X_{(ij)}$ and $Z_{(ij)}$.
The errors $Z_{(ij)}$ and $X_{(ij)}$ can be described as chains $E$ and $E^*$
on the original and dual lattices. The endpoints of the error chains $%
\partial E$ and $\partial E^*$ can be detected by applications of star and
plaquette operators due to anti-commutation of adjacent errors with
operators. From the knowledge of endpoints $\partial E$ and $\partial E^*$
without the homology class of the error chains, error syndrome, we infer the
most likely homology class of error chains, while considering any reasonable
choices. Since $E^{\prime }=E+C$ and ${E^{\prime }}^*=E^*+C^*$, where $C$
and $C^*$ are the contractible loops on both of the lattices, are in
equivalent class with the error chains, the probability for the homology
class $\bar{E}$ and $\bar{E^*}$ of the error chains can be written as \cite%
{Dennis} 
\begin{equation}
P(\bar{E},\bar{E}^*|\partial E,\partial E^*) = P(\bar{E},\bar{E}%
^*)/\sum_{i}P_{D_i}(\bar{E},\bar{E}^*),
\end{equation}
where $P(\bar{E},\bar{E}^*)\propto \sum_{C,C^*}\prod_{\langle
ij\rangle}\exp(K\tau^{E}_{ij}\tau^{C}_{ij}+K\tau^{E^*}_{ij}\tau^{C^*}_{ij})$
for the uncorrelated case. The summation is taken over all the possibilities
of $C$ and $C^*$, and the product is over all the edges. The parameter $K$
stands for the importance/preference to choose the inferred error chain. The
quantity in the denominator $P_{D_i}(\bar{E},\bar{E^*})$ denotes the
probability with the different homology class specified by the logical
operators $D_i$ ($i=1,2,\cdots,2^4$). We here use $\tau^{E}_{ij}$ to
represent the inferred error chains, which takes $\pm 1$ ($\tau^{E}_{ij}<0$,
when $(ij)\in E$), and also for $E^*$, $C$ and $C^*$. The loop constraints $%
\prod_{(ij)}\tau_{ij}^C=1$ and $\prod_{(ij)}\tau_{ij}^{C^*}=1$ allow us to
use another expression by the Ising variables $\tau_{(ij)}^C=\sigma_i\sigma_j
$, and $\tau_{(ij)}^{C^*}=\sigma^*_i\sigma^*_j$ for each lattice on the
torus. By use of these expressions, we can find that $P(\bar{E},\bar{E^*})$
is written as square of the partition function of the $\pm J$ Ising model 
\begin{equation}
P(\bar{E},\bar{E^*}) \propto \sum_{\sigma,\sigma^*}\prod_{\langle ij \rangle}%
\mathrm{e}^{K(\tau^{E}_{ij} \sigma_i\sigma_j+
\tau^{E^*}_{ij}\sigma^*_i\sigma^*_j)}.  \label{PF1}
\end{equation}
where $\tau^{E}_{ij}$ and $\tau^{E^*}_{ij}$ are the signs of the quenched
random couplings in context of spin glasses. When we set $K=K_{\mathrm{ind.}}
$, where $\exp(2K_{\mathrm{ind.}})=(1-p)/p$ (Nishimori line), the inference
of the error chains is an optimal recovery procedure to identify the most
likely homology class \cite{HNbook}. Each of the quenched random couplings
follows the distribution function of the error chains $P(E,E^*)=\prod_{%
\langle ij \rangle}P(\tau^{E}_{ij})P(\tau^{E^*}_{ij})$ for the uncorrelated
case, where 
\begin{eqnarray}
P(\tau^{E}_{ij}) &=& (1-p) \delta_{\tau^{E}_{ij},1} + p
\delta_{\tau^{E}_{ij},-1}.
\end{eqnarray}
Similarly, we can evaluate the probability $P(\bar{E},\bar{E^*})$ for the
homology class of the error chains $\bar{E}$ and $\bar{E^*}$ for the
depolarizing channel case as 
\begin{equation}
P(\bar{E},\bar{E^*}) \propto \prod_{\langle ij \rangle}\mathrm{e}^{K\tau^{%
\bar{E}}_{ij}+K\tau^{\bar{E}^*}_{ij}+K\tau^{\bar{E}}_{ij}\tau^{\bar{E}%
^*}_{ij}},
\end{equation}
where we set the parameter $K=K_{\mathrm{dep.}}$ as on the Nishimori line $%
\exp(4K_{\mathrm{dep.}})=3(1-p)/p$. This is written in terms of the
partition function of the eight-vertex model with quenched random
interaction \cite{AHMMH} 
\begin{equation}
Z_{\mathrm{dep.}} = \sum_{\sigma,\sigma^*}\prod_{\langle ij\rangle}\mathrm{e}%
^{K(\tau^{E}_{ij}\sigma_i\sigma_j+\tau^{E^*}_{ij}\sigma^*_i\sigma_j^*+%
\tau^{E}_{ij}\tau^{E^*}_{ij}\sigma_i\sigma_j\sigma^*_i\sigma^*_j)},
\label{PF2}
\end{equation}
where $\tau^{E}_{ij}$ and $\tau^{E^*}_{ij}$ follow the distribution function
through $P(E,E^*)=\prod_{\langle ij \rangle}P_{\mathrm{dep.}%
}(\tau^{E}_{ij},\tau^{E^*}_{ij})$ as $P_{\mathrm{dep.}}(1,1)=1-p$ while $P_{%
\mathrm{dep.}}(1,-1)=P_{\mathrm{dep.}}(-1,1)=P_{\mathrm{dep.}}(-1,-1)=p/3$.
We emphasize that, if we tune the probability function appropriately, we can
apply our analysis as shown below to inhomogenius case with $p_X \neq p_Y
\neq p_Z$.

In context of the statistical physics, $D_i$ represents the domain wall. In
the low-temperature region implying a small $p$, the order of the degrees of
freedom suppresses the fluctuation of the domain wall. The cost for free
energy difference due to the domain wall diverges as $\sum_{E,E^*}P(E,E^*)P(%
\bar{E},\bar{E}^*|\partial E,\partial E^*)\to 1$ for $L \to \infty$. This
means that we can infer the equivalent class with the original error chains.
On the other hand, in the high-temperature region, the cost vanishes and $%
\sum_{E,E^*}P(E,E^*)P(\bar{E},\bar{E}^*|\partial E,\partial E^*)\to 1/16$.
This implies that the failure of the recovery occurs at the critical point.
Therefore the location of the critical point on the Nishimori line, the
multicritical point, identifies the optimal error threshold.

\begin{figure}[tbp]
\begin{center}
\includegraphics[width=60mm]{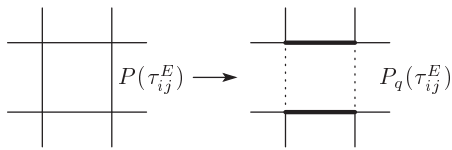}
\end{center}
\caption{Reconstruction of the damaged lattice by use of the weight-zero and
irregular weight. The dashed line denotes the lack of qubits and weight-zero
edge. The bold line expresses the irregular weight edges after the
reconstruction. In this case, $n=2$. }
\label{fig1}
\end{figure}

\textit{Loss of qubits and bond dilution}--- Loss of qubits on the lattice
implies the modification of the stabilizers as well as the logical
operators. However we can reform a complete set of stabilizers even on the
damaged lattice due to loss of qubits following the proposed scheme in Ref. 
\cite{Loss1}. The effect of lost qubits appears in the pattern of error
chains $E$ and $E^*$, and their weight for the probability, which degrades
the performance of error-correcting code. To infer the most likely homology
class based on the knowledge of the error chains on the damaged lattice, we
reconstruct the original lattice by assigning of weight-zero edges on the
lost qubits and irregular weight edges $p^{\prime }$ adjacent to the lost
qubits as $1-2p^{\prime n}$, where $n$ is the number of the shared qubits in
adjacent edges as in Fig. \ref{fig1}. The weight-zero edges imply that we
need to consider a diluted version of the original spin glass system as in
Eqs. (\ref{PF1}) and (\ref{PF2}). It can be achieved by a simple
modification of the distribution function for $\tau^{E}_{ij}$ and $%
\tau^{E^*}_{ij}$ into, for the uncorrelated case, $P^{q}(\tau^E_{ij}) =
(1-q)P(\tau^E_{ij})+q\delta_{\tau^E_{ij},0}$ and $P^q(\tau^{E^*}_{ij})$,
where $q$ denotes the ratio of loss of qubits. Similarly, for the
depolarizing channel case, $P^{q}_{\mathrm{dep.}}(\tau^E_{ij},%
\tau^{E^*}_{ij}) = (1-q)P_{\mathrm{dep.}}(\tau^{E}_{ij},\tau^{E^*}_{ij})+q%
\delta_{\tau^{E}_{ij}\tau^{E^*}_{ij},0}$. In addition, we have to take into
account effects of irregular weight edges $p^{\prime }$ adjacent to the lost
qubits as carefully discussed in Ref. \cite{Loss2}. The effect can be
described by highly correlated distribution function depending on the
pattern of the lost qubits, although we omit its detailed expression. 

\textit{Duality analysis for spin glasses}--- Analyses to clarify the
critical phenomena in finite-dimensional spin glasses are intractable in
general. However a recent development in the spin glass theory enables us to
estimate the precise value of the special critical point on the Nishimori
line, which corresponds to the optimal error threshold \cite%
{NN,MNN,ONB,Ohzeki}. The method as shown below is based on the duality,
which can identify the location of the critical point especially on
two-dimensional spin systems \cite{WuWang}. Let us review the simple pure
Ising model case at first. The duality is a symmetry argument by considering
the low and high-temperature expansions of the partition function $%
Z=\sum_{\sigma_i}\prod_{\langle ij \rangle}\exp(K\sigma_i\sigma_j)$. The
painful calculation of both expansions can be replaced by a simple
manipulation with the binary Fourier transformation for the local part of
the Boltzmann factor, namely edge Boltzmann factor $x_0 = \exp(K)$ and $x_1
= \exp(-K)$ \cite{WuWang}. The low-temperature expansion can be expressed by 
$x_0$ and $x_1$. On the other hand, the high-temperature expansion is given
by the binary Fourier transformation $x_0^*=(x_0+x_1)/\sqrt{2}$ and $%
x_1^*=(x_0-x_1)/\sqrt{2}$. We use this fact and find a double expression of
the partition function as 
\begin{equation}
\{x_0(K)\}^{2L^2}z(u_1(K)) = \{x^*_0(K)\}^{2L^2}z(u^*_1(K)),
\end{equation}
where $z$ is the normalized partition function $z(u_1)=Z/\{x_0(K)\}^{2L^2}$
and $z(u^*_1)=Z/\{x^*_0(K)\}^{2L^2}$. We here define $u_1(K) = x_1(K)/x_0(K)
= \exp(-2K)$ and $u_1^*(K) = x^*_1(K)/x^*_0(K) =\tanh K$. The well-known
duality relation $\exp(-2K^*)=\tanh K$ is given by rewriting $u^*_1(K)$ by $%
u_1(K^*)$, which implies a transformation of the temperature. Then the
principal Boltzmann factors $x_0(K)$ and $x_0^*(K)$ with edge spins parallel
holds $x_0(K_c)=x_0^*(K_c)$ at the critical point $\exp(-2K_c)=\tanh K_c$.

\begin{figure}[tbp]
\begin{center}
\includegraphics[width=60mm]{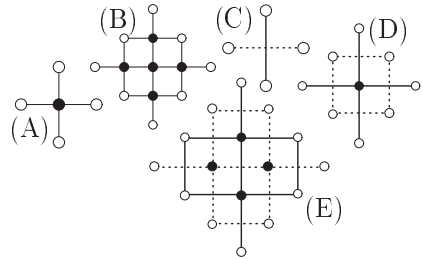}
\end{center}
\caption{Clusters for the uncorrelated and depolarizing channel cases. 
The cluster Boltzmann factor is defined without omitting many body interactions generated after
the decimation of the black spins on the original lattice. 
The dotted line denotes the dual lattice for $\protect\sigma_i^*$ and $\protect\tau^{E^*}_{(ij)}$.}
\label{fig2}
\end{figure}

We employ the replica method, which is often used in theoretical studies on
spin glasses, in order to generalize the duality analysis to spin glasses 
\cite{NN,MNN}. Let us consider the duality for the replicated partition
function as $[Z_{\mathrm{ind.}}^n]$ and $[Z_{\mathrm{dep.}}^n]$ simply $[Z^n]
$, where $[\cdots]$ is the configurational average for the quenched
randomness according to the distribution functions. The multiple ($2^n$)
Fourier transformation again leads us to the double expression of the
replicated partition function as 
\begin{eqnarray}
& & \{x_0(q,K)\}^{2L^2}z(u_1(q,K),u_2(q,K),\cdots)  \notag \\
& & = \{x^*_0(q,K)\}^{2L^2}z(u^*_1(q,K),u^*_2(q,K),\cdots),
\end{eqnarray}
where the subscript of $u_k$ and $u_k^*$ stands for the number of
anti-parallel pair among $n$ replicas on each edge. Unfortunately we cannot
replace $u^*_k(q,K)$ by $u_k(q^*,K^*)$ as the pure case, since the
replicated partition function is multivariable. Nevertheless we can estimate
the precise location of the critical point even for spin glasses by
considering a wider range of the local part of the Boltzmann factor given
after the summation of the internal spins. For instance, in the case on the
square lattice, we define the cluster Boltzmann factor $x_k^{cl.}$, where
the subscript $k$ denotes the configuration of the edge (white-colored)
spins as in Fig. \ref{fig2}. 
We set the equation to lead the location of the critical point as, inspired by the
case without quenched randomness, $x_0(K)=x^*_0(K)$ \cite{NN,MNN,ONB,Ohzeki}%
, 
\begin{equation}
x_0^{\rm cl.}(q,K) = x_0^{\rm cl.*}(q,K).  \label{MCP}
\end{equation}
The equality even without use of the cluster can give the precise solutions of the
critical point for the multicritical point of $\pm J$ Ising model $q=0$ as $%
p_c=0.1100$ \cite{NN,MNN}. Although the above method is not exact, if we
increase the size of the used cluster, we can systematically approach the
exact solution for the critical points of the $\pm J$ Ising model in the
higher temperature region than the Nishimori line \cite{ONB,Ohzeki}.

\begin{figure}[tbp]
\begin{center}
\includegraphics[width=65mm]{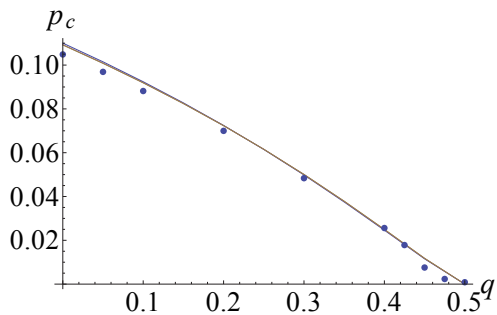}
\end{center}
\caption{(Color online) Results for the uncorrelated case. The dots stand
for numerical data given in Ref. \protect\cite{Loss1}, where the error bars
are suppressed in this scale. The curves almost overlapped in this scale
express our results by the duality.}
\label{fig3}
\end{figure}

\textit{Results}--- In the present study, we consider two clusters as A and B as well as a single edge for the uncorrelated case and a single crossing edges C and two clusters D and E for the depolarizing case as in Fig. \ref{fig2}. 
We show several estimations given by Eq. (\ref{MCP}) for the uncorrelated
case in Table. \ref{table1}. Although, for the uncorrelated case by B cluster, we have considered the highly correlated distribution function by
taking into account the effect of the irregular weight, the results have not
been changed from those by a simple distribution function $P_q(E)$, which is
the same as one for the bond-diluted spin glass. As discussed in Ref. \cite{Loss2}, the highly correlation between the loss of qubits is found to
emerge as a finite-size effect in the numerical investigation (for $q \ge 0.45$). 
Such the complicated effect does not spoil our analysis. 
All the results for any $q$ does not show drastic changes dependently on the size of the used cluster. 
It means that our analyses are enough correct to capture the accurate locations of the optimal error threshold. 
The optimal error thresholds indicate the upper bounds for error threshold by any heuristic methods. 
As shown in Fig. \ref{fig3}, we compare our results with the inference by use of the matching algorithm namely, the ground state as $K\to \infty$ \cite{Loss1}, in which we denote the error thresholds as $p_c^0$. 
We confirm that the heuristic matching algorithm of inference gives $p_c\approx p_c^0$, presumably $p_c = p_c^0$ for large $q$. 
We also give several results for the depolarizing channel case in Table. \ref{table2}. 
Similarly to the case without loss of qubits ($q=0$) as have been reported in Ref. \cite{AHMMH}, the depolarizing channel is more resilient than the
uncorrelated case even with loss of qubits. 
For comparison, let us take an earlier study on an error recovery procedure for the depolarizing channel in Ref. \cite{HM}. 
Our result implies that there is still possibility to improve the performance of such a constructive procedure. 
\begin{table}[htbp]
\begin{center}
\begin{tabular}{lcccc}
\hline
$q$ & $p_c$ & $p_c$ (A) & $p_c$ (B) & $p^0_c$\cite{Loss1} \\ 
\hline
$0.00$ & $0.11003$ & $0.10928$ & $0.10918$ & $0.10486$ \\ 
$0.10$ & $0.09240$ & $0.09196$ & $0.09189$ & $0.08816$ \\ 
$0.20$ & $0.07245$ & $0.07235$ & $0.07233$ & $0.06997$ \\ 
$0.30$ & $0.04984$ & $0.05004$ & $0.05009$ & $0.04836$ \\ 
$0.40$ & $0.02462$ & $0.02492$ & $0.02500$ & $0.02561$ \\ 
$0.45$ & $0.01155$ & $0.01174$ & $0.01179$ & $0.00757$ \\ \hline
\end{tabular}%
\end{center}
\caption{Comparison of the approximations by the clusters A, B, and C for the
uncorrelated case and by a heuristic method \protect\cite{Loss1}. 
We add the improved result given by Ref.\cite{Loss2} to the above list $p_c^{(0)} = 0.1065$. }
\label{table1}
\end{table}
\begin{table}[htbp]
\begin{center}
\begin{tabular}{lcccc}
\hline
$q$ & $p_c$ (C) & $p_c$ (D) & $p_c$ (E) & $p_c$\cite{HM} \\ \hline
$0.00$ & $0.18929$ & $0.18886$ & $0.18852$ & $0.164$ \\ 
$0.10$ & $0.16025$ & $0.15985$ & $0.15960$ & -- \\ 
$0.20$ & $0.12690$ & $0.12656$ & $0.12641$ & -- \\ 
$0.30$ & $0.08844$ & $0.08819$ & $0.08815$ & -- \\ 
$0.40$ & $0.04454$ & $0.04440$ & $0.04443$ & -- \\ 
$0.45$ & $0.02121$ & $0.02114$ & $0.02117$ &-- \\ \hline
\end{tabular}%
\end{center}
\caption{Results for the depolarizing channel case.}
\label{table2}
\end{table}
All the obtained values are almost stable in the third digits. In a
practical sense, our estimations for error thresholds serve as the reference
values.

\textit{Conclusion}--- We have estimated the error thresholds for the
surface code with loss of qubits, via a finite-dimensional spin glass theory, for both of the uncorrelated and
depolarizing channel cases, and shown more resilience of the depolarizing
channel even with loss of qubits.

In the sense of study on spin glass, the comparison between the error
thresholds $p_c^0$ by a suboptimal method corresponding to the inference in
the ground state \cite{Loss1} and optimal ones $p_c$ shows a fascinating
feature of the phase boundary of the $\pm J$ Ising model as $p_c \approx
p_c^0$ \cite{HNbook}. The future study will be desired for solving the
remaining problem on a realm of spin glasses: $p_c = p_c^0$ or not.

\textit{Acknowledgement}--- The author acknowledges fruitful discussions
with and numerical data in Ref. \cite{Loss1} from Thomas Stace and
Sean Barret. He also thanks hospitality in Rome University during this work.
This work was partially supported by MEXT in Japan, Grant-in-Aid for Young
Scientists (B) No.20740218.


\end{document}